\documentstyle[aps,preprint,epsf]{revtex}

\begin{document}
\draft

\title{Connection between Adam-Gibbs Theory\\ and Spatially Heterogeneous
Dynamics}

\author{Nicolas Giovambattista$^1$, Sergey V. Buldyrev$^1$,
Francis W. Starr$^2$ and H. Eugene Stanley$^1$}

\address{$^1$Center for Polymer Studies and Department of Physics\\
  Boston University, Boston, MA 02215 USA}
\address{$^2$Polymers Division and Center for Theoretical and
  Computational Materials Science\\ National Institute of Standards and
  Technology, Gaithersburg, MD 20899 USA} 

\date{September 13, 2002}
\maketitle

\begin{abstract}
We investigate the spatially heterogeneous dynamics in the SPC/E model
 of water by using
molecular dynamics simulations.  We relate the average mass $n^*$ of
mobile particle clusters to the diffusion constant and the
configurational entropy. Hence, $n^*$ can be interpreted as the mass of
the ``cooperatively rearranging regions'' that form the basis of the
 Adam-Gibbs theory of the dynamics of supercooled liquids. Finally, we examine the time and temperature dependence of these
transient clusters.

\end{abstract}



\bigskip\bigskip

\noindent
More than thirty five years ago Adam and Gibbs (AG) proposed a theory to
describe the dynamics of supercooled liquids\cite{adams,debene}. In their
approach they suggest that the system changes its configuration by the
motion of independent ``cooperatively rearranging regions'' (CRR). Their
main result is that the diffusion constant $D$ is related to the
temperature $T$ and the configurational entropy of the system $S_{\rm
conf}$ by 
\begin{mathletters}
\begin{eqnarray}
D \propto \exp (-A/TS_{\rm conf}).
\end{eqnarray}
In the thermodynamic limit,
they interpreted $S_{\rm conf}$ as $k_B\log W_c$, where $W_c$ is the
number of configurations accessible by the system and $k_B$ is the
Boltzmann constant. More recently, $W_c$ has been interpreted as the
number of basins in the potential energy surface (PES) accessible to the
system in equilibrium, facilitating direct calculation of $S_{\rm conf}$
by computer simulations\cite{sconfMath}. The AG prediction has been
tested and appears to be valid across a wide spectrum of liquids
\cite{AntonioNature,sastry}.  The AG theory also hypothesizes a relation
between $S_{\rm conf}$ and the characteristic mass $z$ of the CRR.  However, the CRR
are not precisely defined in the theory, and in the absence of a
definition it has been challenging to test this aspect of AG theory.
 
More recently, computer simulations on simple systems (such as
Lennard-Jones mixtures) have shown that particles of high mobility tend
to form clusters, and the concept of spatially heterogeneous dynamics
 (SHD) is evolving
\cite{kobDonati,donatiDouglas,donatiGlotzer,DH1,DH2,science2D,cicerone,moreExpDH,science3D}.
Sets of neighboring particles move with enhanced or diminished mobility
relative to the average on a time scale intermediate between ballistic
and diffusive motion.  While there has been interest in the possible
relation between clusters obtained from a SHD analysis and the CRR of
the AG theory,
 a link between the quantitative SHD methods and the AG predictions
has not been found.

Here we show that on the time scale where SHD is prominent, the average
cluster size $n^\ast$ can be related to the mass $z$ of CRR, thus
connecting the quantitative SHD analysis to the qualitative approach of
AG.  Our results are based on molecular dynamics simulations of the
extended simple point charge (SPC/E) model of water \cite{spce}.  We
simulate $N=1728$ molecules at fixed density $\rho = 1.0$~g/cm$^3$ and a
range of $T$ from 200~K to 260~K (at 10~K intervals). For each $T$, we
run two independent simulations to improve statistics.
 We find that $D$ can be fit with 
\begin{eqnarray}
D\sim(T-T_{MCT})^\gamma
\label{gamma}
\end{eqnarray}
\end{mathletters}
using the values for the mode coupling temperature $T_{\mbox{\scriptsize
MCT}}=193$~K and the diffusivity exponent $\gamma=2.80$ reported in
Ref.~\cite{francislong}.  

To facilitate comparison with previous work, we use the same approach to
define SHD clusters as that employed to study a Lennard-Jones (LJ)
mixture \cite{donatiGlotzer} and experiments on colloids
\cite{science3D}.  We define the mobility of a molecule at a given time
$t_0$ as the maximum displacement of the oxygen atom in the interval
$[t_0,t_0+\Delta t]$.  Following \cite{donatiGlotzer}, we calculate the
self part of the time-dependent van Hove correlation function\cite{McDonald}
$G_s(r,t)$ at $t=t^*$, the time at which the
non-Gaussian parameter
\begin{eqnarray}
\alpha_2(t) \equiv \frac{3}{5} \langle r^4(t)
\rangle / \langle r^2(t) \rangle^2-1
\end{eqnarray}
 has a maximum \cite{francescoAlpha}.
  We fit $G_s(r,t)$ with a Gaussian approximation
$G_0(r)$ and define $r^*$ as the second intersection between these
distributions.  We find $r^*$ is in the range $0.2-0.25$~nm for all
$T$. We focus on the fraction $\phi$ of ``mobile'' molecules given by
$\phi \equiv \int_{r^*}^\infty 4\pi r^2 G_s(r,t^*) dr$, i.e. the
average fraction of molecules with a displacement larger than $r^*$ in
the interval $t^*$.  Depending on $T$, we find $6\% < \phi < 8\%$.  For
simplicity we fix $\phi=7\%$ for all $T$.  Similar values of $\phi$ were
found in atomic systems~ \cite{donatiDouglas,donatiGlotzer,science3D}
and in polymer melts \cite{francisPol}.  Finally, we define a cluster of
mobile molecules at each interval $\Delta t$ as those mobile molecules
whose nearest neighbor oxygen-oxygen distance at time $t_0$ is less than
the first minima of the oxygen-oxygen radial distribution function
\cite{footnotePablo,footnotedoo}.

\begin{figure}[htb]

\narrowtext 

\centerline{
\hbox {
  \vspace*{0.5cm}  
  \epsfxsize=7cm
  \epsfbox{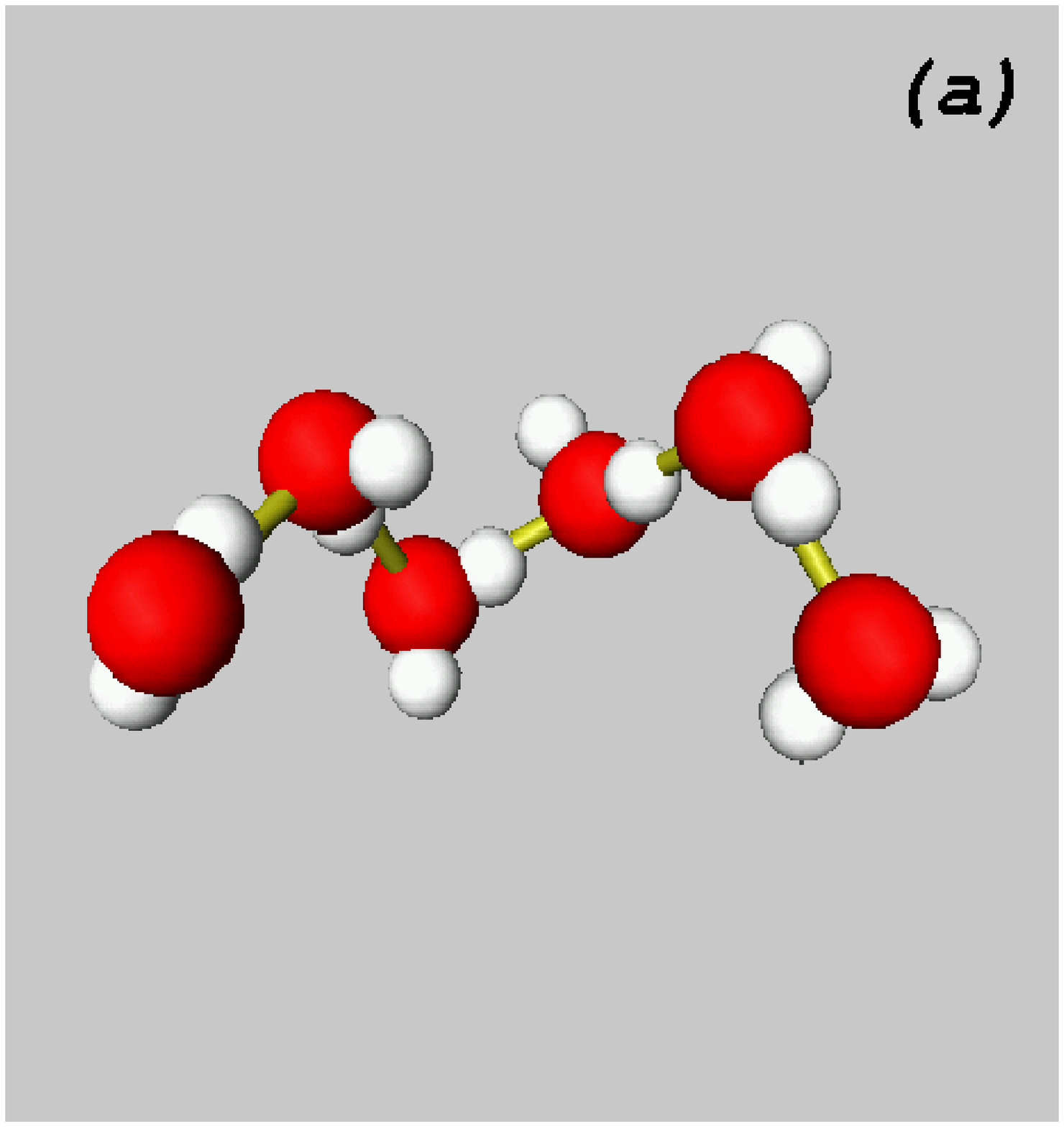}
  \hspace*{0.1cm}
  \epsfxsize=7cm
  \epsfbox{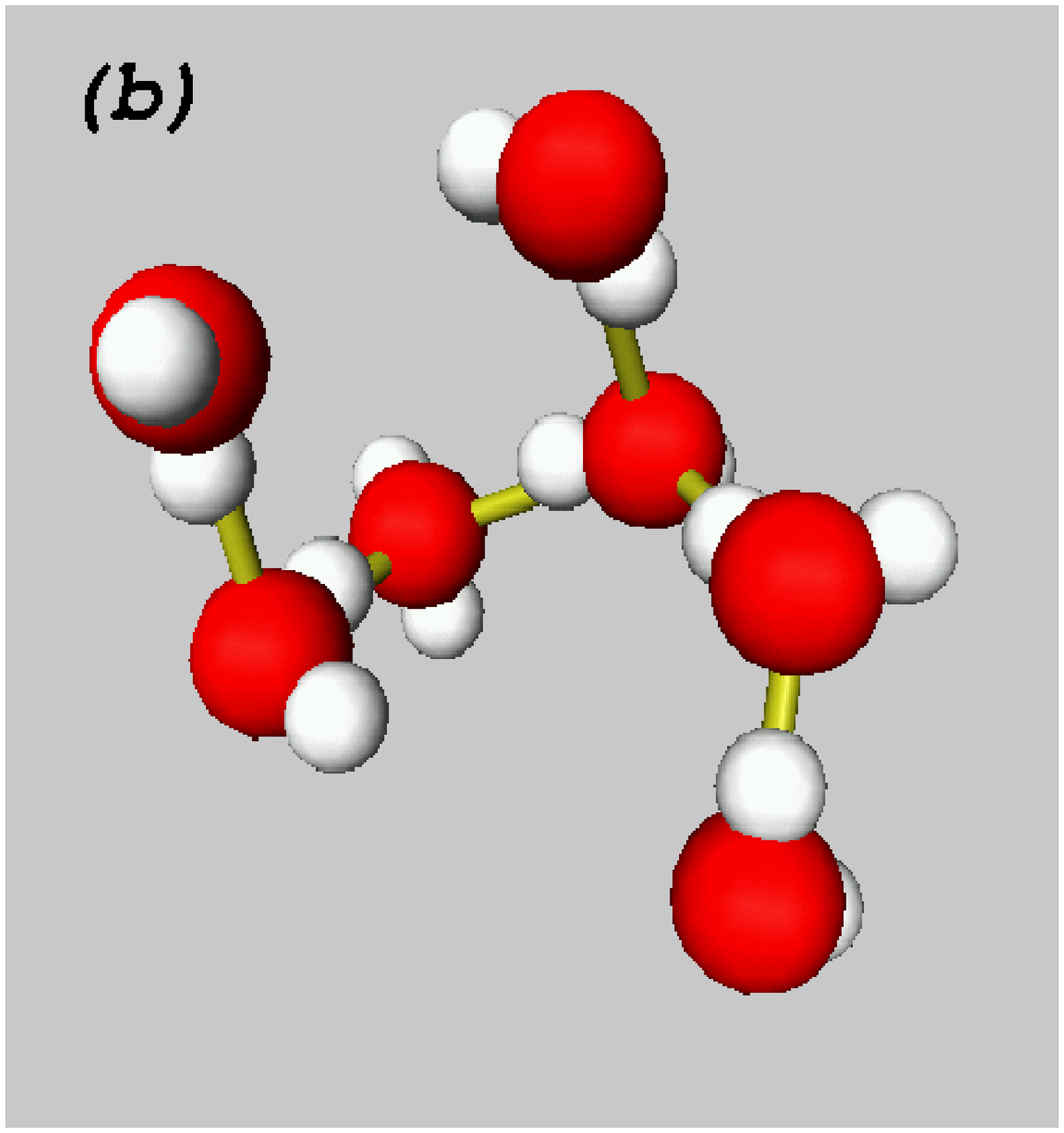}
}}
\centerline{
\hbox {
  \vspace*{0.5cm}  
  \epsfxsize=7cm
  \epsfbox{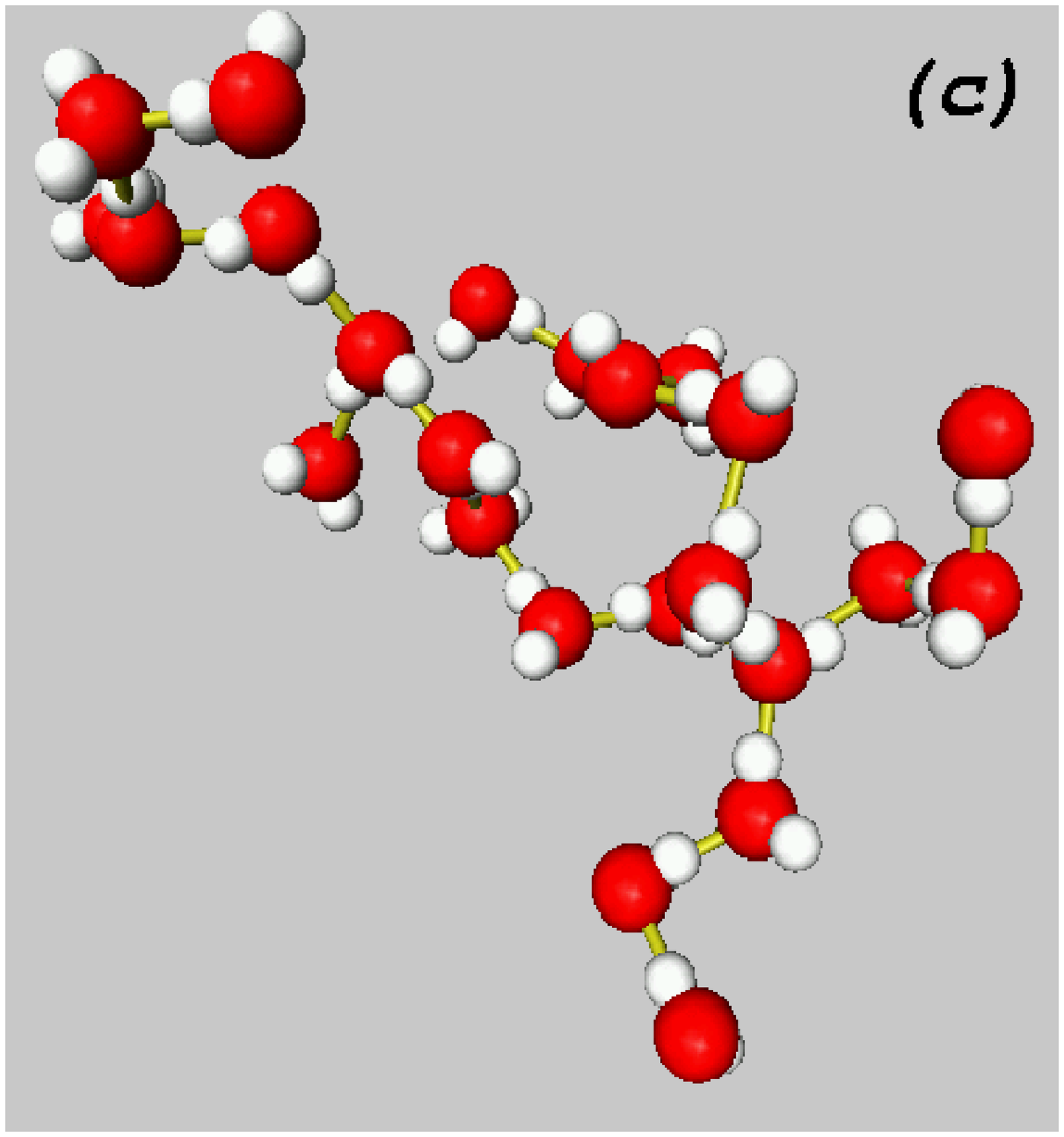}
  \hspace*{0.1cm}
  \epsfxsize=7cm
  \epsfbox{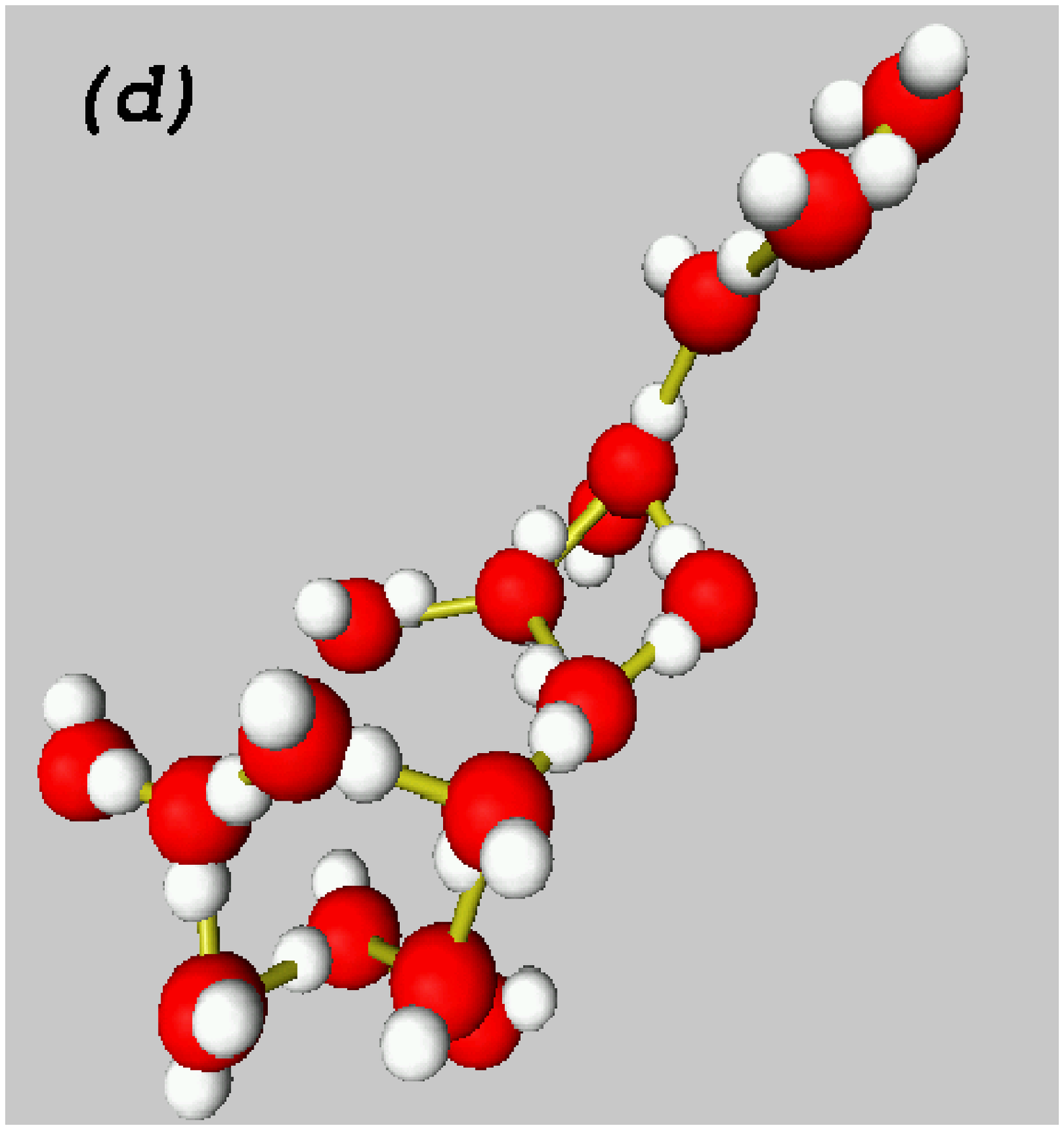}
}}

\vspace*{0.5cm}

\caption{Four clusters of mobile molecules found at $T=210$~K defined with an
  observation time $\Delta t=t^*$.  Tubes connect neighboring molecules
 whose oxygen-oxygen distance is less than $0.315$~nm, the
 first minimum in the oxygen-oxygen radial distribution function.
Small clusters can be either (a) string-like or (b) non-string-like, showing
  branching points (molecules with more than two neighbors).
(c,d) Larger clusters exhibit more complicated structures. Clusters can
  usually be decomposed in substrings, although this can be complicated
  by the presence of loops.
}
\label{snapshotPrl}
\end{figure}

In Fig.~\ref{snapshotPrl} we present four snapshots of mobile particle
clusters at $T=210$~K for $\Delta t=t^*$.  
particles follow each other in a roughly linearly %
fashion\cite{donatiDouglas,science3D}.  In LJ systems
\cite{donatiDouglas}, mono atomic liquids \cite{A-sharon} and polymers
\cite{B-sharon} complex clusters are composed of more elementary
``strings'', where particles follow each other in a roughly linearly
fashion.  For water, small clusters are indeed string-like (e.g.
Fig.~\ref{snapshotPrl}(a)), but the molecules conform to the hydrogen
bond geometry, and hence the clusters are generally less linear than
clusters found in LJ systems.  Like in the simpler systems, clusters
become less string-like as their size increases
\cite{donatiDouglas,C-sharon}, and the fraction of branching
points---molecules with more than two neighbors
(Fig.~\ref{snapshotPrl}(b))---increases with increasing cluster size.
Indeed, the larger the cluster, the more complicated is its
structure---becoming increasingly ramified (Fig.~\ref{snapshotPrl}(c))
and even exhibiting loops
(Fig.~\ref{snapshotPrl}(d))~\cite{string-note}. Hence it appears that
the basic features of SHD found for models of simple liquids
extend to the more complex molecular liquid, water.

To relate SHD to the AG approach, we calculate the average cluster mass
$\langle n(\Delta t) \rangle$ for each $T$.  In the AG approach to
dynamics, the CRR are characterized by the number of particles $z$ and
configurational entropy $s_{\rm conf}(z)$ of the CRR; AG argue that
$z=Ns_{\rm conf}(z)/S_{\rm conf}$.  Motivated by the recent finding that
the average instantaneous cluster mass scales inversely with the entropy
in a model of living polymerization \cite{livingPol}, we use $n^* \equiv
\langle n(t^*) \rangle$ as a measure of $z$, since at $t^*$ correlations
are very pronounced and $\langle n(t) \rangle$ is nearly maximal
\cite{footnote-tt*}. We find a linear relationship between $n^*$ and
$1/S_{\rm conf}$ (Fig.~2(a)),
\begin{eqnarray}
n^* -1 \propto \frac{1}{S_{\rm conf}}.
\label{result}
\end{eqnarray}
This suggests that $n^*-1$ can be understood as a measure of $z$ and provides
a quantitative connection \cite{footnoteConnection} between SHD clusters
and the AG approach. It is necessary to subtract one from $n^*$ to
obtain direct proportionality, implying that a cluster of unit size does
not correspond to a CRR\cite{donatiGlotzer}.  Equation~(\ref{result}) provides a clear link
between a cluster property, $n^*$, and a property of the PES, $S_{\rm conf}$.
 Furthermore,
given that $S_{\rm conf}$ and the diffusion constant $D$ have been
previously related \cite{AntonioNature}, it follows that
\begin{eqnarray}
D \sim e^{-A(n^* -1)/T}.
\end{eqnarray}
Our results in Fig.~2(b) confirm this expectation.

We next address the question of how SHD clusters in water depend on the
observation time $\Delta t$. We focus on the number average $\langle
n(\Delta t) \rangle$ and $\langle n(\Delta t) \rangle_{w}=\langle
n^2(\Delta t) \rangle/\langle n(\Delta t) \rangle$; $\langle n(\Delta t)
\rangle_{w}$, the weight average cluster size, is the average size
of a cluster to which a randomly chosen molecule belongs.
Figure~\ref{SnM} shows $\langle n(\Delta t) \rangle$ and $\langle
n(\Delta t) \rangle_{w}$ for $T=210$~K. To eliminate the random
contribution, we normalize $\langle n(\Delta t) \rangle_{w}$ by $\langle
n_r \rangle_{w}$, the weight average cluster size for $\phi N$ randomly chosen
molecules. For comparison, we also include the non-Gaussian parameter
$\alpha_2(\Delta t)$ and the mean-squared displacement $\langle
r^2(\Delta t)\rangle$ (Fig.~\ref{SnM}(a)) which displays the three
characteristic time regimes, ballistic, cage and diffusive.

The behavior of $\langle n(\Delta t) \rangle_{w}/\langle n_r
\rangle_{w}$ is analogous to that for polymer systems \cite{francisPol}, with
the exception that there is a clear increase in $\langle n(\Delta t)
\rangle_{w}/\langle n_r \rangle_{w}$ at the time scale on which
molecules go from the ballistic to the cage regime. This additional
feature is likely due to strong correlations in the vibrational motion
of the first-neighbor molecules, owing to the presence of hydrogen
bonds.  In Fig.~\ref{SnM}(c) we show $\langle n(\Delta t)
\rangle_{w}/\langle n_r \rangle_{w}$ for all $T$.  For $T \leq 240$~K,
the maximum in $\langle n(\Delta t) \rangle_{w}/\langle n_r \rangle_{w}$
increases in magnitude and shifts to larger time scales with decreasing
$T$. The plateau at the crossover from the ballistic regime is nearly
$T$-independent, as expected since the mean collision time is nearly
$T$-independent.  For $T \geq 250$~K, the maximum and the plateau merge,
and hence it is not possible to separately distinguish these features.

We find that $t^*$ is slightly larger than 
$t_{\mbox{\scriptsize max}}$, the time where the maximum of
$\langle n(\Delta t) \rangle_{w}$ occurs.
Both characteristic times correspond to the
late-$\beta$/early-$\alpha$ time regime of the mode coupling theory
(MCT). We find (Fig.~\ref{tmax})
\begin{mathletters}
\begin{eqnarray}
t^* \sim (T-T_{\mbox{\scriptsize MCT}})^{-\delta},~~[\delta=2.7 \pm 0.1]
\end{eqnarray}
and
\begin{eqnarray}
t_{\mbox{\scriptsize max}} \sim (T-T_{\mbox{\scriptsize MCT}})^{-x},~~
[x=2.7 \pm 0.1]
\label{expx}
\end{eqnarray}
\end{mathletters}
For the LJ polymer melt, simulations show that $x=1/2a$, where $a$ is the
scaling exponent predicted by MCT for the $\beta$
time scale, suggesting that $t_{\mbox{\scriptsize max}}$ may be a
measure of the ``elusive'' $\beta$ relaxation time scale
\cite{francisPol}.  MCT predicts that knowing $\gamma$, defined in
Eq.~(\ref{gamma}), is sufficient for determining $a$
\cite{mct}. If $t_{\mbox{\scriptsize
max}}$ were a measure of the $\beta$-time scale, then we would expect
$a=0.28$ \cite{francislong}, and hence $x$ would be equal to $1.78$. From
Eq.~(\ref{expx}), we see that $x>1.78$, so
 $t_{\mbox{\scriptsize max}}$ does \textit{not} provide a measure
 of the $\beta$ time scale.
 Additional tests of the temperature scaling of
$t_{\mbox{\scriptsize max}}$ for other liquids are needed to determine
the range of liquids for which $t_{\mbox{\scriptsize max}}$ can be
considered a measure of the $\beta$ time scale.

In summary, the relation we find between $n^*$ and $S_{\rm conf}$,
 Eq.~(\ref{result}), provides a link
between SHD and properties of the PES\cite{schroder}. In the context of
AG theory, our findings support the interesting possibility  
 that $n^*$ is a measure of the size of the cooperatively
rearranging regions. Furthermore, our simulations show that
SHD in water are qualitatively similar to those found in LJ systems, but
the cluster shapes are strongly influenced by the geometry of the
hydrogen bond network.

\bigskip\bigskip

\noindent
We thank J.F. Douglas, Y. Gebremichael, S.C. Glotzer, T.G. Keyes,
 S. Mossa and F. Sciortino for fruitful discussions,
and S. Kamath and S. Kumar for sharing their ideas on 
the relation between the cluster mass and
$S_{\rm conf}$ for a lattice model of a dense
polymer melt.
This work was supported by the NSF Chemistry Program.

\newpage

\begin{figure}[htb]
\narrowtext \centerline{
\hbox {
  \vspace*{0.5cm}  
  \epsfxsize=13cm
  \epsfbox{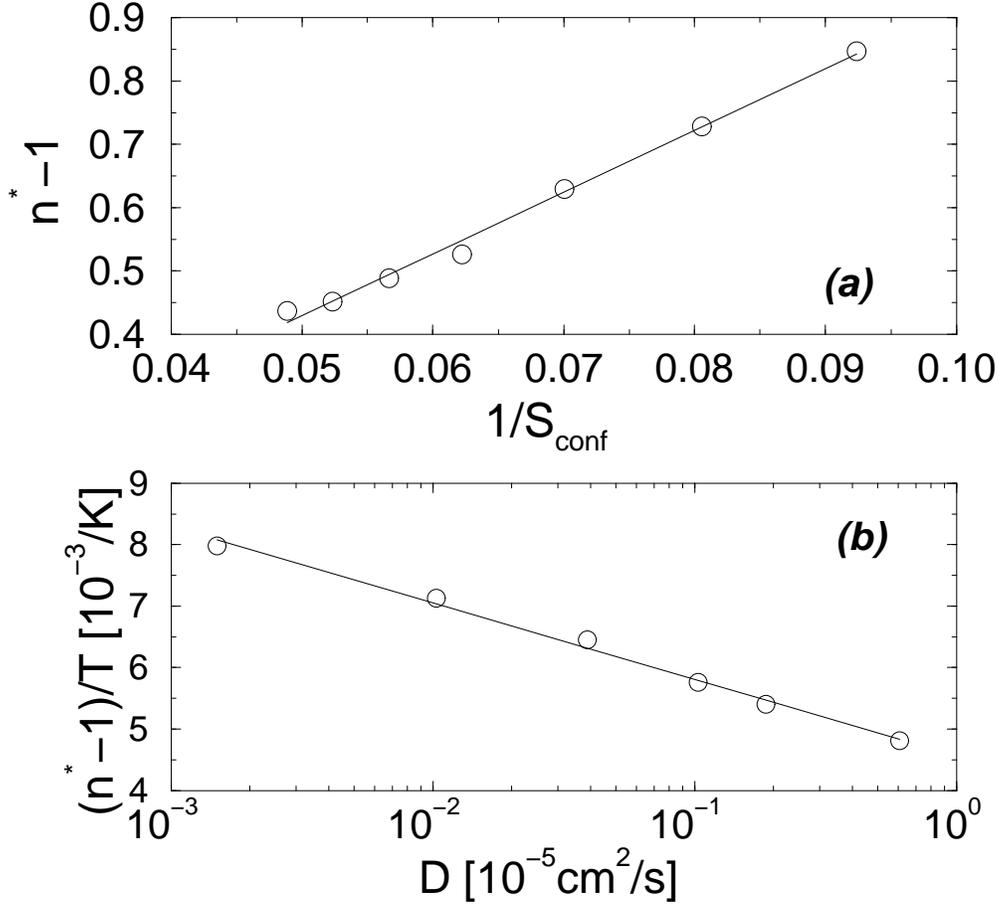}
  \hspace*{0.3cm}
  }
}

\vspace*{0.5cm}

\caption{(a) The average cluster size $n^*$ is proportional to the
inverse of the configurational entropy $S_{\rm conf}$ suggesting that
$n^*-1$ can be used as a measure of the size of the cooperatively
rearranging regions hypothesized by Adam and Gibbs. (b) Log-linear plot
of $(n^*-1)/T$ as a function of the diffusion constant D. The AG
prediction $D \sim \exp\left(A/TS_{\rm conf} \right)$ implies that $\log
D \sim (n^*-1)/T$. This relationship holds for almost three decades in
$D$.  }
\label{S-1overSconf}   
\end{figure}

\begin{figure}[htb]
\narrowtext \centerline{
\hbox {
  \vspace*{0.5cm}  
  \epsfxsize=10cm
  \epsfbox{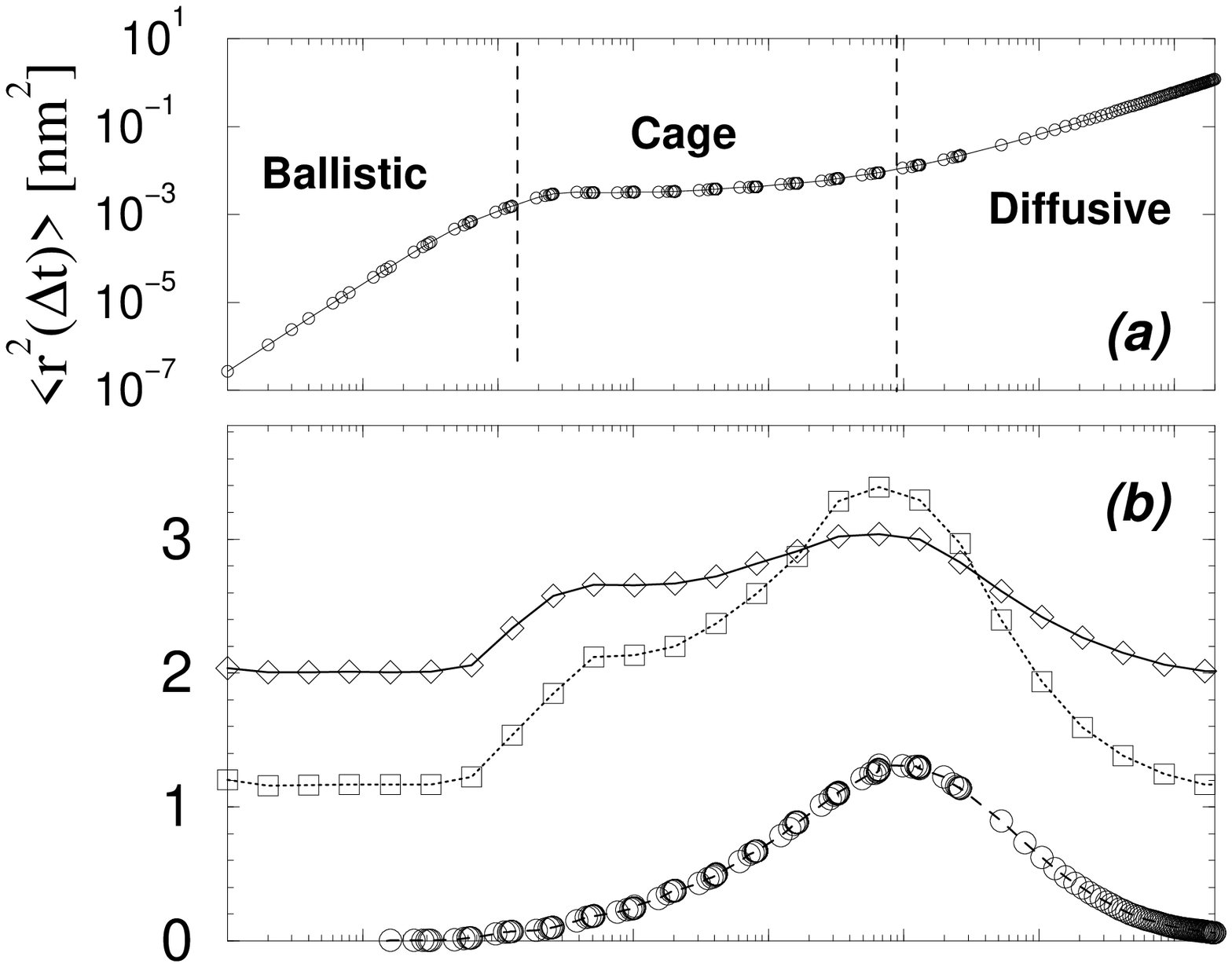}
  \hspace*{0.3cm}
  }
}

\centerline{
\hbox {
  \vspace*{0.5cm}  
  \epsfxsize=10cm
  \epsfbox{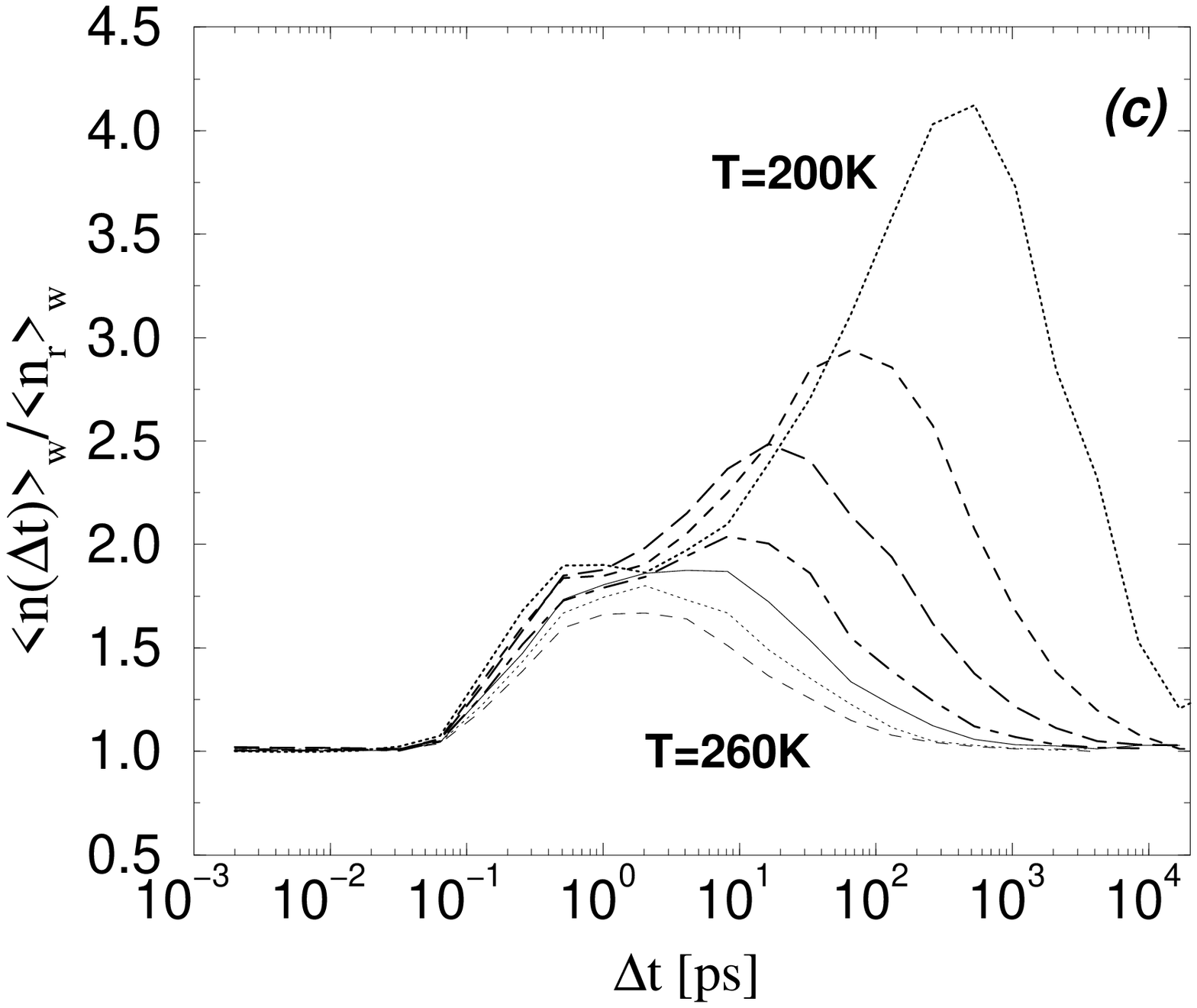}
  \hspace*{0.3cm}
  }
   }

\caption{(a) Mean square displacement $\langle r^2(\Delta t) \rangle$ at
$T=210$~K showing the ballistic, cage and diffusive regimes (separated
by dotted lines).  (b) Average number of molecules $\langle n(\Delta
t\rangle)$ ($\Diamond$) and normalized weight cluster size $\langle n(\Delta
t) \rangle_{w}$ ($\Box$). The behavior of all quantities correlate
 with $\langle
r^2(\Delta t)\rangle$. The maxima of $\langle n(\Delta t) \rangle_{w}$
and $\langle n(\Delta t\rangle)$ occur at times slightly smaller than
the time for the maximum in $\alpha_2(\Delta t)$ ($\bigcirc$),
 the non-Gaussian
parameter. (c) Weight cluster size $\langle n(\Delta t)
\rangle_{w}/\langle n_r \rangle_{w}$ for temperatures ranging from
$200$~K to $260$~K in intervals of $10$~K. Note the $T$-independent
plateau at the crossover from ballistic motion.  }
\label{SnM}
\end{figure}

\begin{figure}[htb]
\narrowtext \centerline{
\hbox {
  \vspace*{0.5cm}  
  \epsfxsize=13cm
  \epsfbox{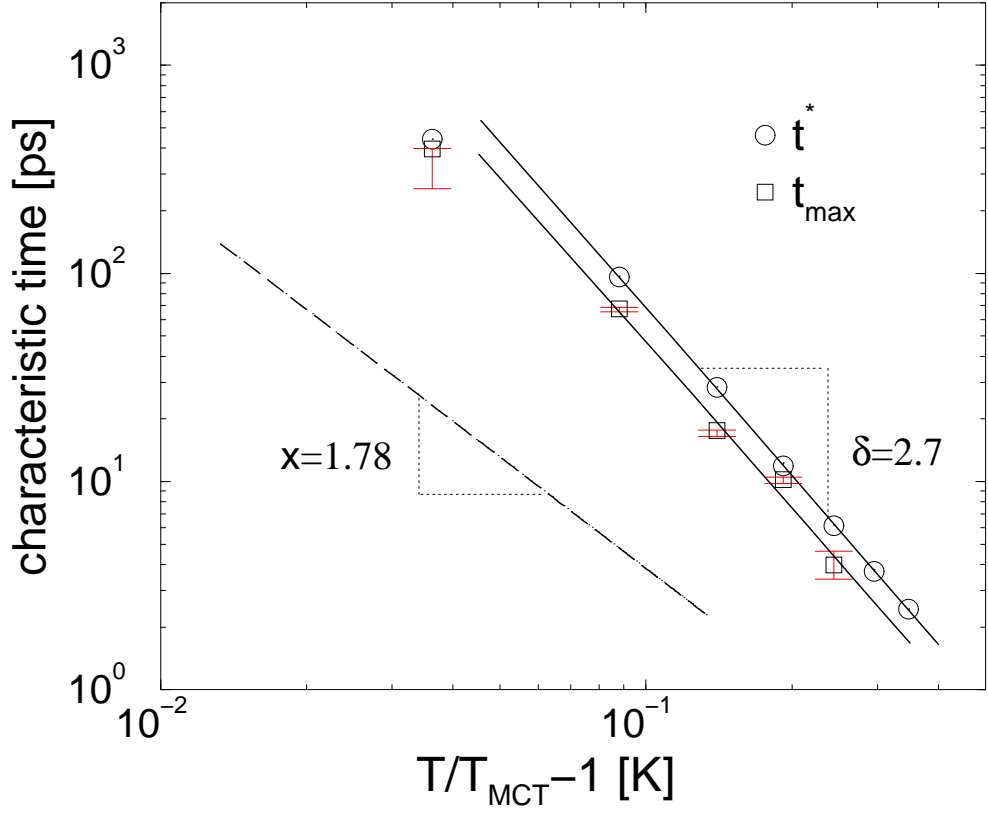}
  \hspace*{0.3cm}
  }
}

\vspace*{0.5cm}

\caption{Temperature dependence of $t^*$ and $t_{\mbox{\scriptsize
max}}$, the times at which the maxima of the non-Gaussian parameter and
the weight average cluster size occur. We find $t^* \sim
(T/T_{MCT}-1)^\delta$ and $t_{\mbox{\scriptsize max}} \sim
(T/T_{MCT}-1)^x$ where $x = \delta = 2.7 \pm 0.1$.  The expected value
of the exponent $x$, following arguments in \protect\cite{francisPol},
is $1.78$ (long dashed line). The values of $t^*$ and
$t_{\mbox{\scriptsize max}}$ for $T=200$~K were not included in the fits
because $200$~K is too close to $T_{MCT}=193$~K (deviations from MCT are
known to occur as $T \rightarrow T_{MCT}$).  }

 \label{tmax} \end{figure}

\vspace*{1.0cm}


\begin{references}
\bibitem{string-note} Note that clusters in
Fig.~\protect\ref{snapshotPrl}(c) can be decomposed in substrings, as
found for LJ systems.
\bibitem{footnote-tt*} The maximum of $\langle n(\Delta t) \rangle$
occurs at time slightly before $t^*$. Our conclusions are unaffected by
choosing $n^*$ or the maximum of $\langle n(\Delta t) \rangle$.
\bibitem{footnoteConnection} This connection relies on the assumption
that the $T$ dependence of $s_{\rm conf}(z)$ is weak in comparison to
that of $S_{\rm conf}$, as can be expected since $z \ll N$ and the
configurational entropy is an extensive property.
\end{references}
\end{document}